\documentclass{article}

\usepackage[british]{babel}
\usepackage[useregional]{datetime2}
\DTMlangsetup[en-GB]{showdayofmonth=false}
\usepackage[numbers, sort]{natbib}
\usepackage[letterpaper,top=2cm,bottom=2cm,left=2.5cm,right=2.5cm,marginparwidth=1.75cm]{geometry}
\usepackage{amsmath}
\usepackage{amsfonts}
\usepackage{amssymb}
\usepackage{mathtools}
\usepackage{wrapfig}
\usepackage{graphicx}
\usepackage{tikz}
\usepackage[colorlinks=true, allcolors=blue]{hyperref}
\usepackage[page]{appendix} %

\usepackage{graphicx}
\usepackage{subcaption}
\usepackage{multicol}
\usepackage{lipsum}
\usepackage{float}

\renewcommand{\vec}[1]{\boldsymbol{\mathbf{#1}}}
\renewcommand{\v}{\vec}

\newcommand{\sumiN}{\sum_{i=1}^N}

\makeatletter
\newif\ifnobrackets
\renewcommand\@cite[2]{\ifnobrackets\else[\fi{#1\if@tempswa , #2\fi}\ifnobrackets\else]\fi\nobracketsfalse}
\newcommand\nbcite{\nobracketstrue\cite}
\makeatother
\usepackage{authblk}

\title{\vspace*{0cm}{\textbf{\Large{Rebalancing-versus-Rebalancing: Improving the fidelity of Loss-versus-Rebalancing}}}}
\author{Matthew Willetts}
\author{Christian Harrington}
\affil{QuantAMM.fi}
\begin{document}
\maketitle

\begin{abstract}

Automated Market Makers (AMMs) hold assets and are constantly being rebalanced by external arbitrageurs to match external market prices.
Loss-versus-rebalancing (LVR) is a pivotal metric for measuring how an AMM pool performs for its liquidity providers (LPs) relative to an idealised benchmark where rebalancing is done not via the action of arbitrageurs but instead by trading with a perfect centralised exchange with no fees, spread or slippage.
This renders it an imperfect tool for judging rebalancing efficiency between execution platforms.

We introduce Rebalancing-versus-rebalancing (\emph{RVR}), a higher-fidelity model that better captures the frictions present in centralised rebalancing. 

We perform a battery of experiments comparing managing a portfolio on AMMs vs this new and more realistic centralised exchange benchmark---RVR.
We are also particularly interested in dynamic AMMs that run strategies beyond fixed weight allocations--Temporal Function Market Makers.
This is particularly important for asset managers evaluating execution management systems.
In this paper we simulate more than 1000 different strategies settings as well as testing hundreds of different variations in centralised exchange (CEX) fees, AMM fees \& gas costs.

We find that, under this modeling approach, AMM pools (even with no retail/noise traders) often offer superior execution and rebalancing efficiency compared to centralised rebalancing, for all but the lowest CEX fee levels. 
We also take a simple approach to model noise traders \& find that even a small amount of noise volume increases modeled AMM performance such that CEX rebalancing finds it hard to compete.
This indicates that decentralised AMM-based asset management can offer superior performance and execution management for asset managers looking to rebalance portfolios, offering an alternative use case for dynamic AMMs beyond core liquidity providing.

\end{abstract}
\section{Introduction}

Automated Market Makers (AMMs) are a variety of decentralised exchange (DEX) where a mathematical formula is used on-chain to structure and execute exchanges out of a pool containing a basket of assets.
Liquidity providers (LPs) deposit reserves of assets into the system, which are used as working capital against which the third-party traders can exchange.

AMM pools charge fees on trades, giving a revenue stream to LPs.
These pools can be thought of, however, as a form of asset management.
The mathematical function that defines which trades against pool-held-reserves are accepted \emph{implicitly defines a strategy} for the allocation of value between assets in the pool.

In this form of asset management, the AMM pool reserves \emph{are} the portfolio holdings.
Rebalancing occurs due to the action of arbitrageurs: if the pool holds anything other than its desired allocation it quotes off-market prices.
Trades that take advantage of these off-market prices push the pool towards holding its desired portfolio and reduces the mis-pricing.
In other words, the pool pays arbitrageurs to rebalance.

\paragraph{How can one measure the performance of an AMM pool?}

Early on, impermanent loss (also known as \textit{divergence loss})~\cite{il, boueri2022g3mimpermanentlossdynamics} was used to study AMM performance.
It is the phenomena that,
ignoring fee revenue and for vanilla AMMs, changes in pool reserves (and thus pool value) from changes in market prices (post LP-deposit) lead to a loss of value for LPs compared to if they simply held their initial assets.

Loss-Versus-Rebalancing is a metric that compares the P\&L of an AMM pool versus that of a idealised portfolio that aims to mirror the pool.

In a two token AMM pool we can interpret one token as the risk asset, with reserves $x$, and the other as the risk-off asset, with reserves $y$.
We can then define $P$ as the price of $x$ using $y$ as numeraire.
The pool's value is $V_\mathsf{2token}(t):= P(t)x(t)+y(t)$.
The pool's reserves change over time from the action of traders and arbitrageurs.

LVR introduces a `rebalancing portfolio' with value $V_\mathsf{rebal}(t)$ that aims to match continuously the holdings of an AMM pools' risk asset.
The difference is that the rebalancing portfolio does its rebalancing not by the actions of traders/arbitrageurs but instead by trading externally at the market price perfectly frictionlessly (zero slippage and zero costs).

In continuous time
$V_\mathsf{rebal}(t) = V_\mathsf{2token}(0) + \int_0^t x(P_s)dP_s$
for $t>0$.\footnote{Note that here we are parameterising the strategies holdings $x(\cdot)$ not directly by time but indirectly via the holdings the AMM pool has as function of market prices that themselves vary with time, highlighting that it is changes in market prices that lead to changes in AMM pool holdings.}
LVR is then defined as
\[\mathsf{LVR}(t):= V_\mathsf{rebal}(t) - V_\mathsf{2token}(t).\]

LVR has been extended to handle when constant function market makers charge fees on their trades (called `ARB')~\cite{lvr_arb}.
Fees mean that there is only an arbitrage opportunity if the pool's quoted prices sufficiently deviate from market prices to leave the pool's no-trade (a.k.a. no-arb) region.
Within the setup studied (risk asset prices diffuse under geometric Brownian motion, $V_\mathsf{rebal}(t)$ unchanged, low fees, fast blocks), the presence of CFMM fees decreases LVR by the fraction of time that a arbitrage trade is possible~\cite{lvr_arb}.

\paragraph{Measured performance of fixed weight AMMs } Most AMM pools have static trading functions, where the pool aims to have a constant division of value between assets (for example, a Uniswap V2 pool aims to hold half of its value in each token in the pair).
These strategies are known variously as constant-mix strategies~\cite{dynamic_strats}, constant-weight asset allocation or constant rebalanced portfolios.
These strategies can perform well~\cite{opt_v_naive}, but what both IL and LVR are telling us is that implementing these strategies via an AMM pool is a particularly poor choice of rebalancing method.

So far LVR and its extensions have only been used to compare efficiency of constant-mix strategies that (absent trading fees) tend to perform badly when done by AMMs.

\paragraph{Dynamic AMMs}

Usually an AMM at market equilibrium presents an arbitrage opportunity only when the external market price changes. 
However an AMM can create an arbitrage opportunity by changing the offered price, even if the external market price has remained the same.
The standard geometric mean market maker (G3M) trading function is $\prod_{i=1}^N R_i^{w_i}= k$, where $\sum_{i=1}^N w_i = 1$ and $\forall i\,\, 0\leq w_i<1$,
where $\v R =\{R_i\}$ is the $N$-vector of currently-held reserves of each token and the weights $\v w =\{w_i\}$ control how much value is stored in each token.

There has been recent work around dynamic weight extensions of AMMs where the pool instantiates a dynamic strategy; the strategy dictates whether the pools constituents should be heavily weighted in one constituent and not another.
Pools with these properties are known as Temporal Function Market Makers~\cite{tfmm_litepaper}.
The time-varying trading function that defines a TFMM pool block-to-block is:
\begin{equation}
    \prod_{i=1}^N R_i^{w_i(t)}= k(t), \quad\mathrm{where\,} \sum_{i=1}^N w_i(t) = 1, \,\,\mathrm{and\,\,} \forall i\,\, 0< w_i(t)<1.
    \label{eq:TFMM}
\end{equation}
The value in the pool is simply $V_\mathsf{pool}:=\sumiN R_i(t) p_i(t)$, where $\v p(t)=\{p_i(t)\}$ are the market prices of the $N$ assets in the pool in a chosen numeraire.
Under the action of arbitrageurs, we have $w_i(t) V_\mathsf{pool}(t) \approx R_i(t) p_i(t)$, with equality in the limit of zero fees, zero gas costs and short block times.

If a pool was initially weighted 50:50 BTC/ETH and BTC was increasing in price, a strategy might change $\v w(t)$ over time to rebalance the pool so it holds more BTC, therefore giving LPs the advantage of the price increase.
There are multi-block MEV implications ~\cite{willetts2024multiblockmevopportunities} and optimal trajectories ~\cite{willetts2024optimalrebalancingdynamicamms} to take into consideration within this framework.

\paragraph{TFMM weight change process} At a pool defined interval, say every hour or day, a function is called to update the strategy signal by querying latest price oracle values.
When the new target weights are calculated based on the strategy, weight trajectories are then stored on-chain. 
These on-chain trajectories (for example, increase the pool's BTC weight 0.001\% every block for the next 12 hours and decrease the ETH weight similarly) allow for oracle-independent swap prices using solely the standard AMM pricing formula.
When the price changes enough to to expose an arbitrage opportunity, an external aggressive arbitrageur will take that arbitrage opportunity, rebalancing pool holdings in the desired direction.\footnote{Liquidity bootstrap pools (LBPs) are an example of this principle, however LBPs only change weights in one direction and their weight trajectory is determined on creation, not by hourly or daily oracles. 
}
See~\cite{tfmm_litepaper,quantamm_litepaper} for more detail, including on strategy implementations.

\paragraph{Evaluating dynamic weight AMMs vs centralised trade execution}

TFMMs architecture allow for potentially extremely large pools/porfolios to utilise the vast world of external arbitrageurs bots to rebalance via the change of a handful of on-chain variables---the strategy signal and weight trajectories.
This avoids requiring a multi-venue trade routing stack and avoids off-chain operational costs.
Putting aside the hard to model value add of these benefits, here we wish to attack this question:
\textit{Is this method of execution management more efficient than others?}
LVR is insufficient as a tool as the LVR benchmark (perfect rebalancing) is not a product LPs can access, as its assumptions cannot be met in the real world. 

These considerations leads us to introduce Rebalancing-Versus-Rebalancing (or RVR, pronounced \textit{reever}), where we can model dynamic and well as static strategies.

\section{Rebalancing Versus Rebalancing}

Benchmarks are useful to the extent that they can inform decisions.
While LVR is a useful abstraction, we propose a benchmark that models with higher fidelity how rebalancing against a centralised exchange is done.
Rebalancing a portfolio incurs costs.
Our goal is to capture more closely the range of available approaches that are potentially accessible to LPs.

We aim to provide a benchmark that is equatable to one of the most widely used methods for portfolio execution management: rebalancing using a centralised exchange (CEX).
Here we add modelling of some of the costs associated with CEX trades, while also extending these concepts to multi-asset ($N>2$) portfolios in a simple way.
Given we are now comparing two rebalancing mechanisms, rather than an idealised benchmark and a target rebalancing system, we call the resulting metric Rebalancing-versus-Rebalancing or RVR.

The following table compares the core features of LVR, its fee extended model ARB and RVR:

 \begin{table}[h!]
 \centering
  \begin{tabular}{|| c | c c c ||} 
  \hline
   \textbf{Model Features} & LVR~\cite{lvr} & ARB~\cite{lvr_arb} & RVR (\emph{this paper}) \\ [0.5ex] 
  \hline\hline
    CEX Spread & 0  & 0 & TradFi~\nbcite[\S2.2]{index_book}  \\
    CEX Fees & 0 & 0 & Fee Present  \\
    AMM Fees & 0 & Fee Present & Fee Present  \\
    AMM Gas Cost & 0 & 0 & Fixed Costs  \\
    AMM tokens & 2 & 2 & N  \\ 
   [1ex] 
  \hline
  \end{tabular}
 \end{table}

\subsection{Multi Asset Portfolios}

We extend LVR-style analysis to multi-asset portfolios.
LVR aims to match holdings of the risk asset $x$ in a \emph{pair}, with profit appearing via a greater holding of the risk-off asset $y$ (used as the numeraire).
In a multi-asset portfolio it is not clear how to extend this logic especially if you do not want to single out one asset as being risk-off.
So rather than aiming to match particular reserve amounts, we treat all members of the portfolio equally and have the RVR rebalancing portfolio aim to match the \emph{ratio} of distribution of portfolio value of the pool, the portfolio vector, over time.
See Appendix~\ref{app:multi_asset_lvr} for more discussion.

\subsection{Centralised portfolio modelling}
In RVR we model the fees charged by centralised exchanges and the presence of the spread for the rebalancing portfolio.
We do this following industry standard approaches, as described in~\nbcite{index_book}.

\paragraph{Rebalancing trade}
The centralised rebalancing portfolio has reserves $\v R_\mathsf{cex}(t) =\{R_{\mathsf{cex},i}(t)\}$.
The change in reserves $\v R_\mathsf{cex}(t) - \v R_\mathsf{cex}(t-1)$ needed to rebalance a portfolio is
\[\v R_\mathsf{cex}(t) - \v R_\mathsf{cex}(t-1)=\frac{\v w(t)V_\mathsf{cex}(t)}{\v p(t)} - \frac{\v w(t-1)V_\mathsf{cex}(t-1)}{\v p(t-1)},\]
where $V_\mathsf{cex}(t):=\sumiN R_{\mathsf{cex},i}(t)p_i(t)$ is the value of our CEX-rebalanced portfolio and all division between vectors is done elementwise.
For how the value changes as prices and weights change and the portfolio is rebalanced, we follow the approach of~\nbcite[\S2.2]{index_book}.
This gives us
\[V_\mathsf{cex}(t)=\sumiN R_{\mathsf{cex},i}(t-1)p_i(t) - c(\v R_\mathsf{cex}(t),\v R_\mathsf{cex}(t-1), \v p(t)),\]
where $c(\cdot)$ is the cost of doing the rebalancing trade.
Here we decompose the costs of doing a trade into costs from commission fees and from the presence of the spread: $c(\cdot)=c_\mathsf{fees}(\cdot) + c_\mathsf{spread}(\cdot)$.

We assume that the market provides infinite liquidity at the bid and ask prices, i.e. we maintain the assumption from LVR that there is no slippage.
This means that we do not include market impact here, but we note that in preliminary investigations using standard market impact modeling approaches~\nbcite[Chapter 16]{Grinold} we find that including it makes vanishingly small difference to the results.

\paragraph{Commission fees}

A simple model for commission fees is to charge a fee amount on the outgoing leg of the trade from the CEX.
We will parameterise the fees as $\tau_{\mathsf{cex}}$, ($\tau_{\mathsf{cex}}=1-\gamma_{\mathsf{cex}}$).

We denote the outgoing leg of the trade $\v\Delta$, so
$\Delta_i = \left(R_{\mathsf{cex},i}(t)-R_{\mathsf{cex},i}(t-1)\right)\mathbb{I}_{R_{\mathsf{cex},i}(t)-R_{\mathsf{cex},i}(t-1)>0}$,
where $\mathbb{I}_{\cdot}$ is an indicator function that returns 1 if the expression in its subscript is true, otherwise returning 0.
$c_\mathsf{fees}$ is then simply
\[c_\mathsf{fees}(\v R_\mathsf{cex}(t), \v R_\mathsf{cex}(t-1), \v p(t))=\tau_{\mathsf{cex}}\sumiN p_i(t)\Delta_i.\]

\paragraph{Spread}
We model the rebalancing portfolio as trading via market orders, so buying at the ask and selling at the bid, following~\nbcite[\S2.2]{index_book}.
For given bid-ask spreads $\v s(t)$ (each given as a proportion in the same numeraire as the market prices $\v p(t)$) for the $N$ assets present, $c_\mathsf{spread}$ is simply
\[c_\mathsf{spread}(\v R_\mathsf{cex}(t), \v R_\mathsf{cex}(t-1), \v p(t))=\frac{1}{2}\sumiN p_i(t)s_i(t)\lvert R_{\mathsf{cex},i}(t)-R_{\mathsf{cex},i}(t-1)\rvert.\]

\paragraph{Calculating the trades}
Putting the above together, we have a system of equations from which we want to obtain the post-rebalancing value of the portfolio $V_\mathsf{cex}(t)$ and the trade we took to get us there, $\v R_\mathsf{cex}(t) - \v R_\mathsf{cex}(t-1)$.
The trade you want to do depends on the cost, and the cost depends on the trade.
This problem is not solvable analytically, but using fast numerical solvers we can calculate $V_\mathsf{cex}(t)$ and $\v R_\mathsf{cex}(t) - \v R_\mathsf{cex}(t-1)$ as needed in our simulations.

\subsection{AMM portfolio modelling}

\paragraph{Trade}
AMMs charge fees on their trades.
It is reasonable to take this into account in modelling the performance of AMM pools.
In the first part of this analysis will assume no noise trades at all, just that arbitrageurs trade with the pool when it is in their interest to do so.
Later on in the results section we will make use of a simple model for noise trades.
In these calculations we benefit from recent advances in closed form expressions for the optimal arbitrage trade against G3M pools~\cite{willetts2024closedform}.

Although TFMM pools benefit from being arbed quickly, as soon as the arbitrage opportunity first appears, we apply a 5 block discovery period, which for ETH mainnet amounts to one minute.
This is to make it so that in our simulations we are not relying on the immediate action of arbitrageurs.
Decreasing this discovery period would only increase performance.

\paragraph{Gas Costs}
Standard modelling the no-arbitrage (aka `no-trade') region of a pool---the combinations of quoted prices and market prices for which arbitrageurs cannot gain from trading against the pool
---assumes that the arbitrageur trades where the market value of the trade gives a return $>0$.

One obvious cost that arbitrageurs have to cover is the gas cost of their own transactions.
We require that the arbitrage trade has to produce a profit greater than a threshold USD amount, which we will vary in our analysis.

\subsection{RVR Definition summary}

RVR is thus the value of a strategies portfolio when implemented via an AMM pool minus the value when that strategy is run via a CEX with commission fees and a spread.
In addition then to depending on the prices of the assets over time, this means that RVR further depends on the choice of three \emph{rebalancing parameters}: pool fees $\gamma$, the gas cost arbitrageurs have to pay to trade with a pool, and the commission fees of the CEX $\tau_{\mathsf{cex}}$.

Thus we define RVR as :
\begin{equation}
\mathsf{RVR}(t):=V_\mathsf{pool}(t)-V_\mathsf{cex}(t),
    \label{eq:defn_rvr}
\end{equation}
so $\mathsf{RVR}(t)>0$ means that the pool is outperforming CEX rebalancing.\footnote{Note that this is a change of sign compared to LVR---in LVR the AMM is always beaten by idealised, perfect centralised rebalancing so that \emph{loss} is given as a positive number. Here that would result in a negative number}
Also we are often interested in $V_\mathsf{pool}(t)$ and $V_\mathsf{cex}(t)$ themselves (for example in Fig~\ref{fig:rvr_intro}).
Finally, RVR scaled by initial pool value is often useful, which is also the difference between the pool's returns and the CEX portfolio's returns.
\[\frac{\mathsf{RVR}(t)}{V(0)}=\frac{V_\mathsf{pool}(t) - V_\mathsf{cex}(t)}{V(0)}.\]

\subsection{Benchmark Use Methodology}

To effectively compare rebalancing methodologies one must attempt to keep as many other variables constant and test ranges of variables that cannot be kept constant. 

Aspects that can be kept constant include rebalancing intervals, target market weights, constituent price sources and initial portfolio values.
Here is an example of weight vector $\v w(t)$ over time for a BTC:ETH:DAI strategy.
 \begin{table}[h!]
 \centering
  \begin{tabular}{|| c | c c c c c ||} 
  \hline
   \textbf{Rebalance Method} & T & T+1m & T + 2m & T + 3m & T + 4m \\ [0.5ex] 
  \hline\hline
    AMM (BTC:ETH:DAI) & 30:60:10  & 29:63:8 & 28:65:7 & 30:65:5 & 30:60:10  \\
    RVR (BTC:ETH:DAI) & 30:60:10 & 29:63:8 & 28:65:7 & 30:65:5 & 30:60:10 \\
   [1ex] 
  \hline
  \end{tabular}
  \caption{Example sequence of weights over 5 minutes for a BTC:ETH:DAI strategy. For both an AMM pool implementing this strategy and for the RVR centralised benchmark, the desired portfolio weight vector is the same.}
 \end{table}

Even though sequence of weights are the same for the AMM pool and the RVR benchmark, under the presence of AMM fees and gas costs, the amounts traded may vary between the two.
For example, the weight change from block to block may not be enough for the deviation in the AMM's price to exit the no-trade region, so a larger weight change may need to `build up' for an external arbitrageur to rebalance the AMM portfolio.
These differences in reserves over time are exactly the source of the different P\&L for different strategies.

\newpage
\section{Results}

We focus here on a portfolio with three assets: BTC, ETH and DAI.
We run simulations of pools with these constituents using Binance market data from Jan 2021 to June 2024, 3 and a half years at minute-level resolution.
To make a conclusion regarding general efficiency benefits there are some variables that must be stress tested.
These include type of strategy (therefore different weight change trajectories and overall trade volume patterns) and both CEX and AMM trading costs including fees and gas cost.
By varying both CEX and AMM rebalancing parameters we can study asset management both for dynamic AMMs and constant function AMMs.

At first we model AMM performance with no noise traders present, with the only potential fees being on arbitrageurs rebalancing the pool.

\subsection{Strategy variations}
For dynamic AMMs the other aspect that greatly affect results is the type of portfolio strategy applied.
This could alter rebalancing speed, volume and frequency and therefore drastically change the overall requirements of the execution management. To eliminate strategy bias a large range of strategies must be tested.

For the dynamic TFMM pools we use a momentum strategy~\cite{quantamm_litepaper} where an increasing price for an asset leads to an increasing weight.
The simplest momentum strategy for TFMM pools has two parameters, a memory length (over what time period are we measuring changes in prices) and an aggressiveness parameter (called $k$ which says, given a change in prices, by how much we should change an asset's weight).

Altering pool strategy parameters provides a wide range of rebalancing requirements and variations over the test period. 
This ultimately leads to a range of volume traded and aggressiveness of the trading. 

\begin{figure}[h!]
\centering
\vspace{-1em}
\begin{subfigure}[b]{0.48\textwidth}  
            \centering 
        \includegraphics[width=\textwidth]
        {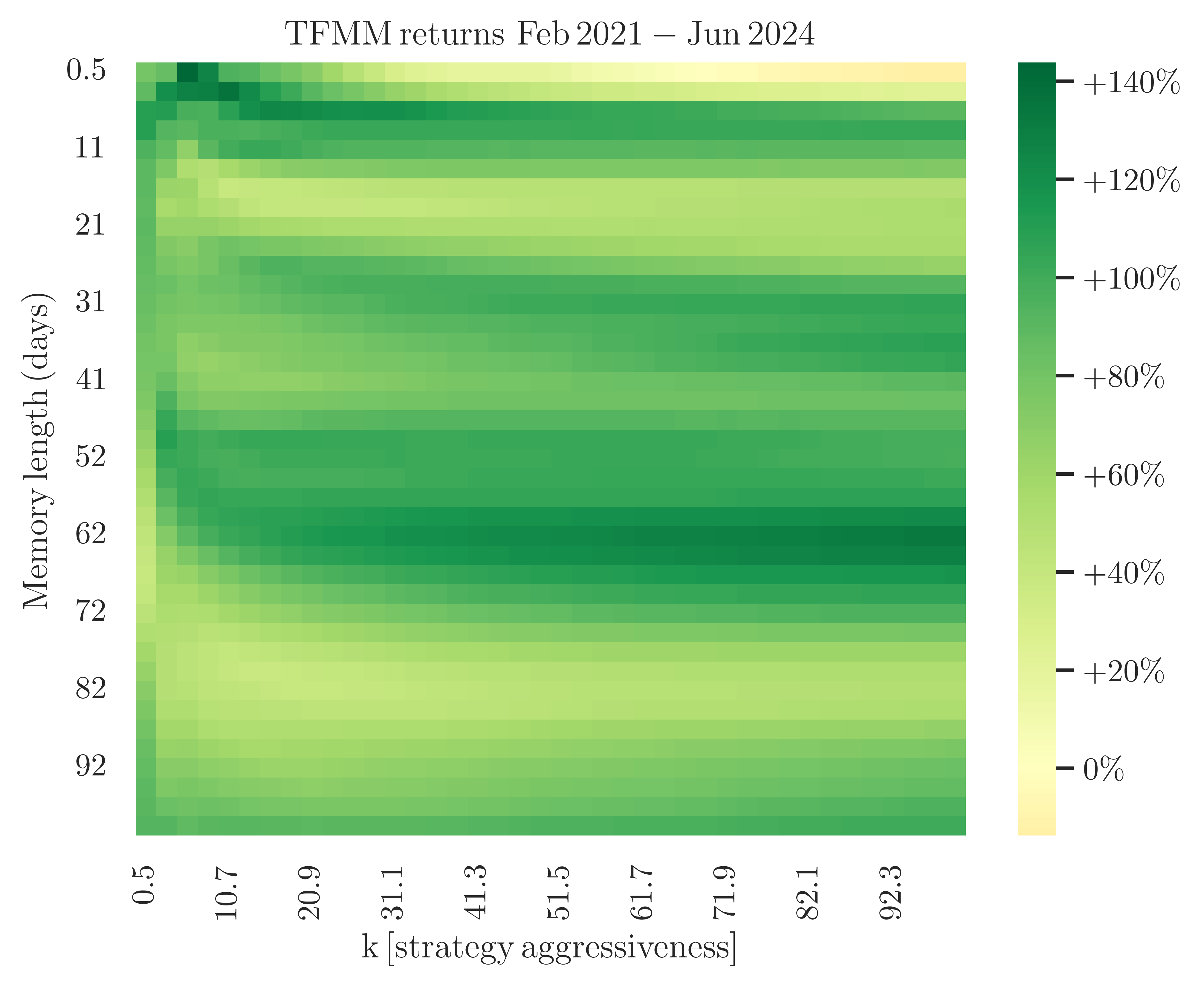}
        \caption[]%
            {{\small Absolute performance }}    
            \label{fig:tfmm_param_returns}
        \end{subfigure}
\begin{subfigure}[b]{0.47\textwidth}
            \centering
        \includegraphics[width=\textwidth]{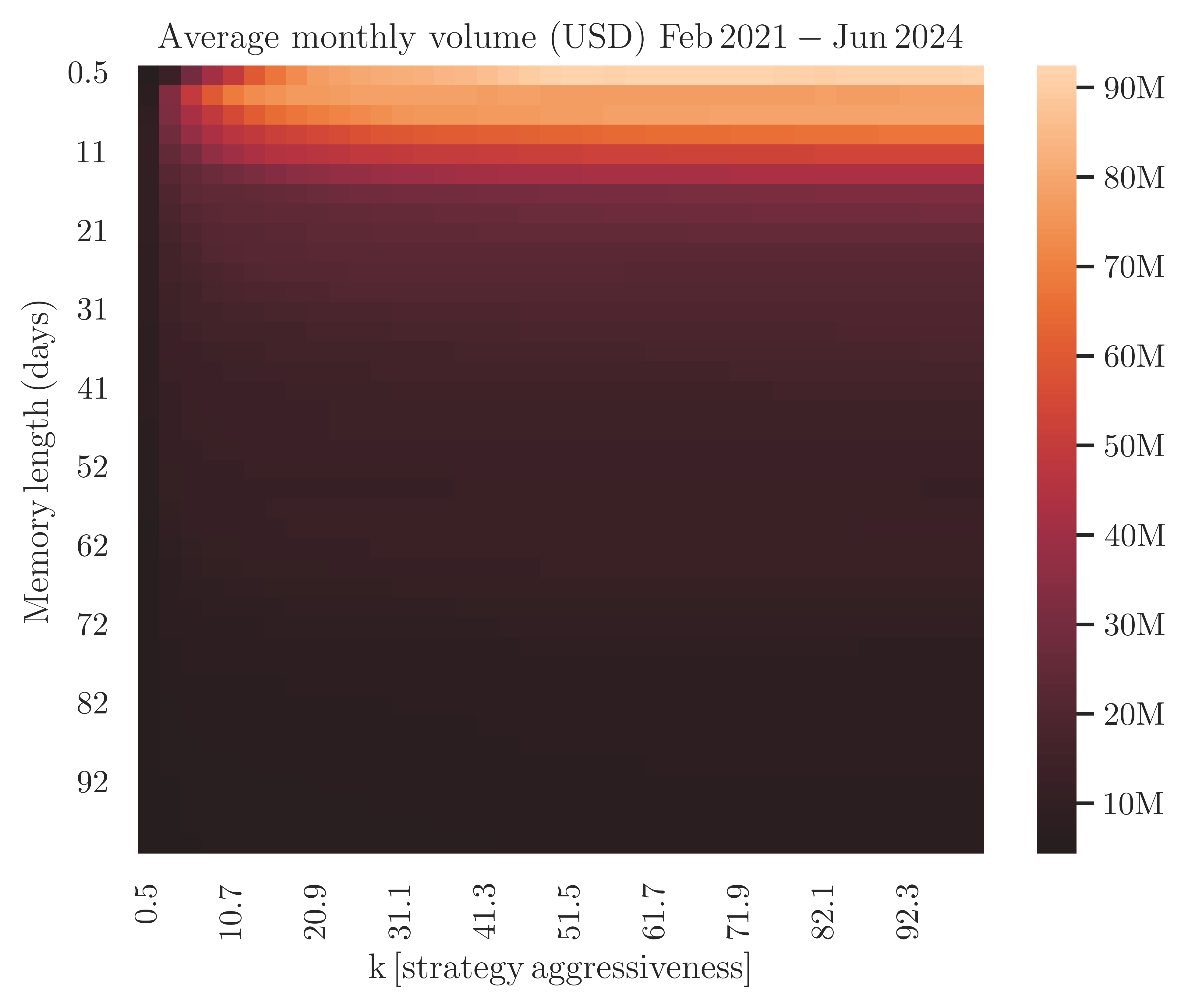}
        
            \caption[]%
            {{\small Average monthly volume}}    
            \label{fig:tfmm_param_rvr}
        \end{subfigure}
     \caption {{Outcomes for 1600 strategy variations. TFMM return difference compared to RVR. a) TFMM pool returns b) average monthly volume (millions USD).$\frac{\mathsf{RVR}(t)}{V(0)}$, RVR scaled by initial value, Bboth
            for the same 1600 variations of momentum strategies running for BTC, ETH, DAI from Feb 2021 to June 2024. 
     These plots come from 40 different memory lengths (measured in days, y-axis) and 40 different levels of agressiveness ($k$, x-axis)}.}    
     \label{fig:rvr_heatmap}
\end{figure}

\paragraph{Heatmaps for 1600 strategies}
Absolute strategy performance can naturally be shown as a heatmap, Fig~\ref{fig:rvr_heatmap}.a), where the colour of each x-y position in the plots are the results for a differently-parameterised momentum strategy.
We also show the average monthly volume for each strategy in Fig~\ref{fig:rvr_heatmap}.b).
These simulations were done with $\tau_\mathsf{cex}=10\mathrm{bps}$, $\gamma=1-0.014$ (so fees of $1.4\%$) and a gas cost of $1$ USD on an initial pool value of $10$ million USD.

\newpage
\paragraph{1600 strategies with different pool fees and gas costs}

How does RVR vary for these strategies if the pool fee, pool trade gas cost vary?
We can hold each constant and vary the other giving us.
We plot these as histograms, Fig~\ref{fig:rvr_hists}, showing RVR for these 1600 different strategies under these variations.

As you see below, for the range of strategies tested regardless of pool fees or gas costs there is a distinct improvement in overall rebalancing efficiency on the RVR benchmark for running these strategies as dynamically weighted AMMs (TFMMs).

\begin{figure}[H]
\centering

\begin{subfigure}[b]{0.7\textwidth}
            \centering
        \includegraphics[width=\textwidth]{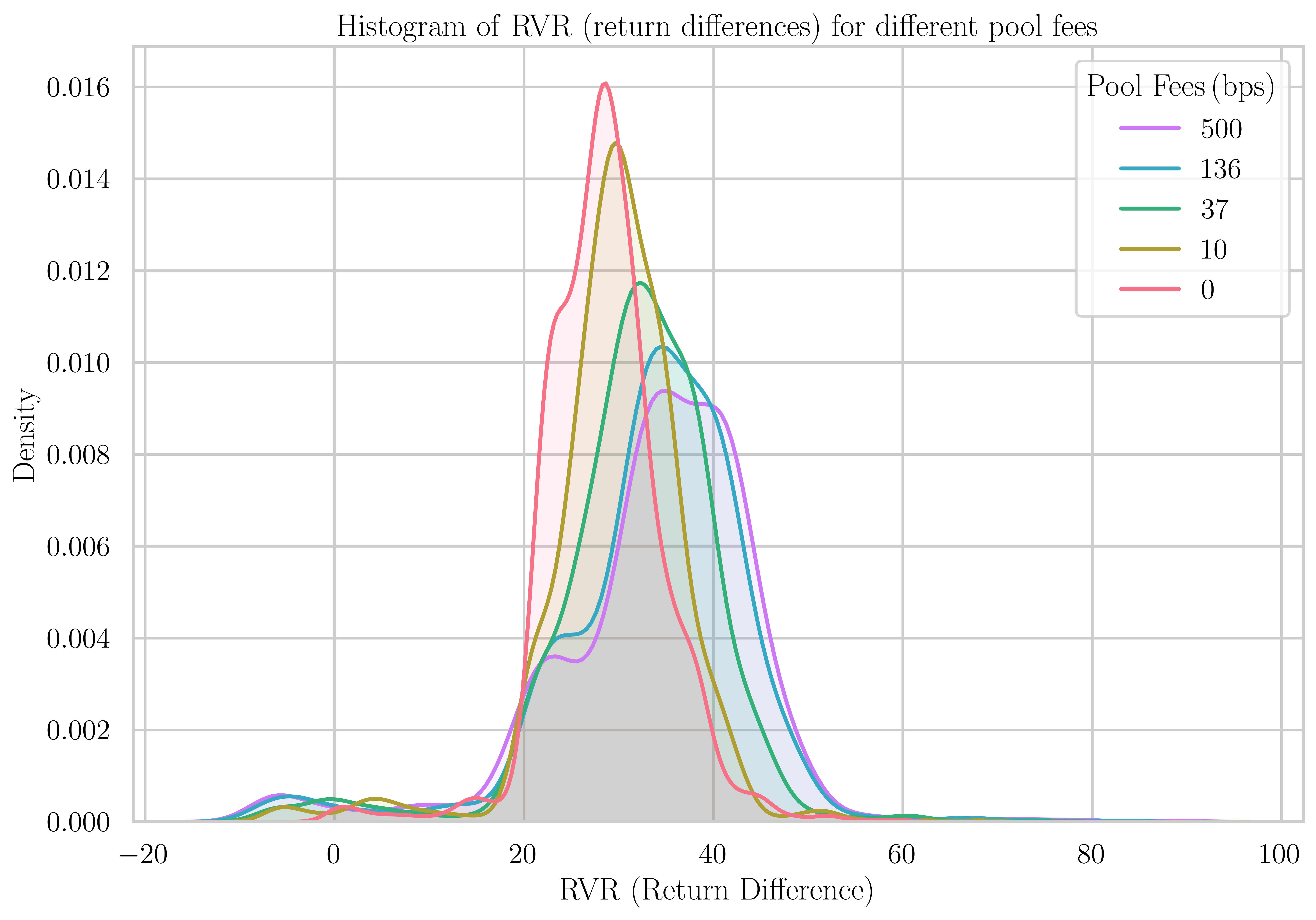}
            \caption[]%
            {{\small RVR when varying pool fees, gas cost = $1$ USD.}}    
            \label{fig:tfmm_hist_varying_fees}
        \end{subfigure}
        \hspace{1em}
        \begin{subfigure}[b]{0.7\textwidth}  
            \centering 
        \includegraphics[width=\textwidth]{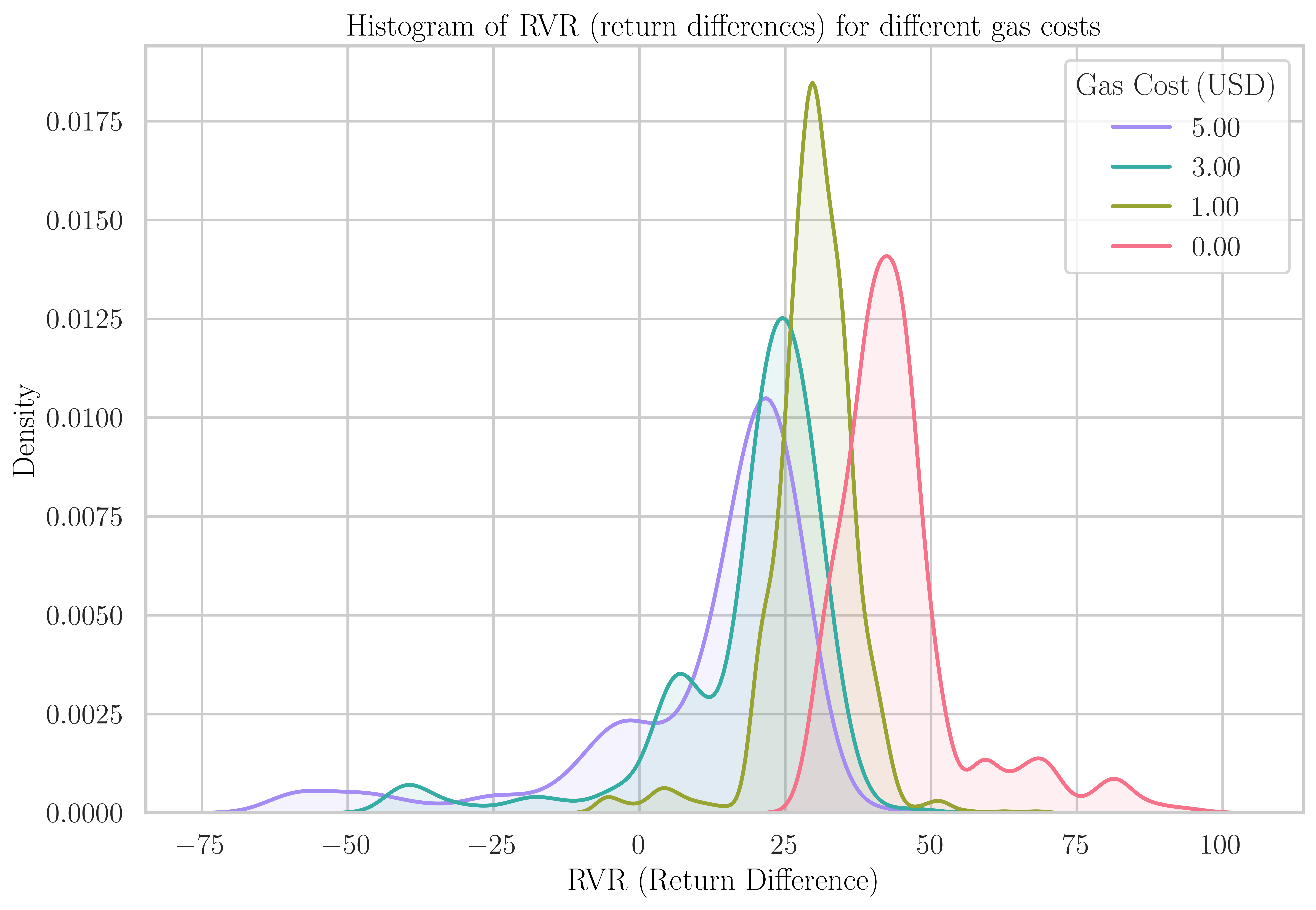}  
        \caption[]%
            {{\small RVR when varying trade gas cost, pool fees = $10$ bps.}}    
            \label{fig:tfmm_hist_varying_gascost}
        \end{subfigure}
            \caption[Network2]%
            {{Each histogram shows density plots showing $\frac{\mathsf{RVR}(t)}{V(0)}$, RVR scaled by initial value, for the 1600 different strategies in Fig~\ref{fig:rvr_heatmap}.
            In a) we hold the gas cost of a trade against the pool constant (at 1USD) but vary the pool fees.
            Higher pool fees tend to increase pool returns and thus RVR, which is intuitive as we compound trading fees.
            In b) we hold the pool fees constant (10bps) and vary the gas cost.
            Lower gas costs increase pool returns and thus RVR, which makes sense as it makes pools easier to arb.
            Both plots show that RVR is generally positive, indicating that running strategies as an AMM pool can have better performance than rebalancing these strategies against a CEX.}    }
            \label{fig:rvr_hists}
\end{figure}
\newpage

\subsection{Results over time}

\begin{figure}[H]
\centering
\vspace{-1em}
\begin{subfigure}[b]{0.45\textwidth}  
            \centering 
        \includegraphics[width=\textwidth]{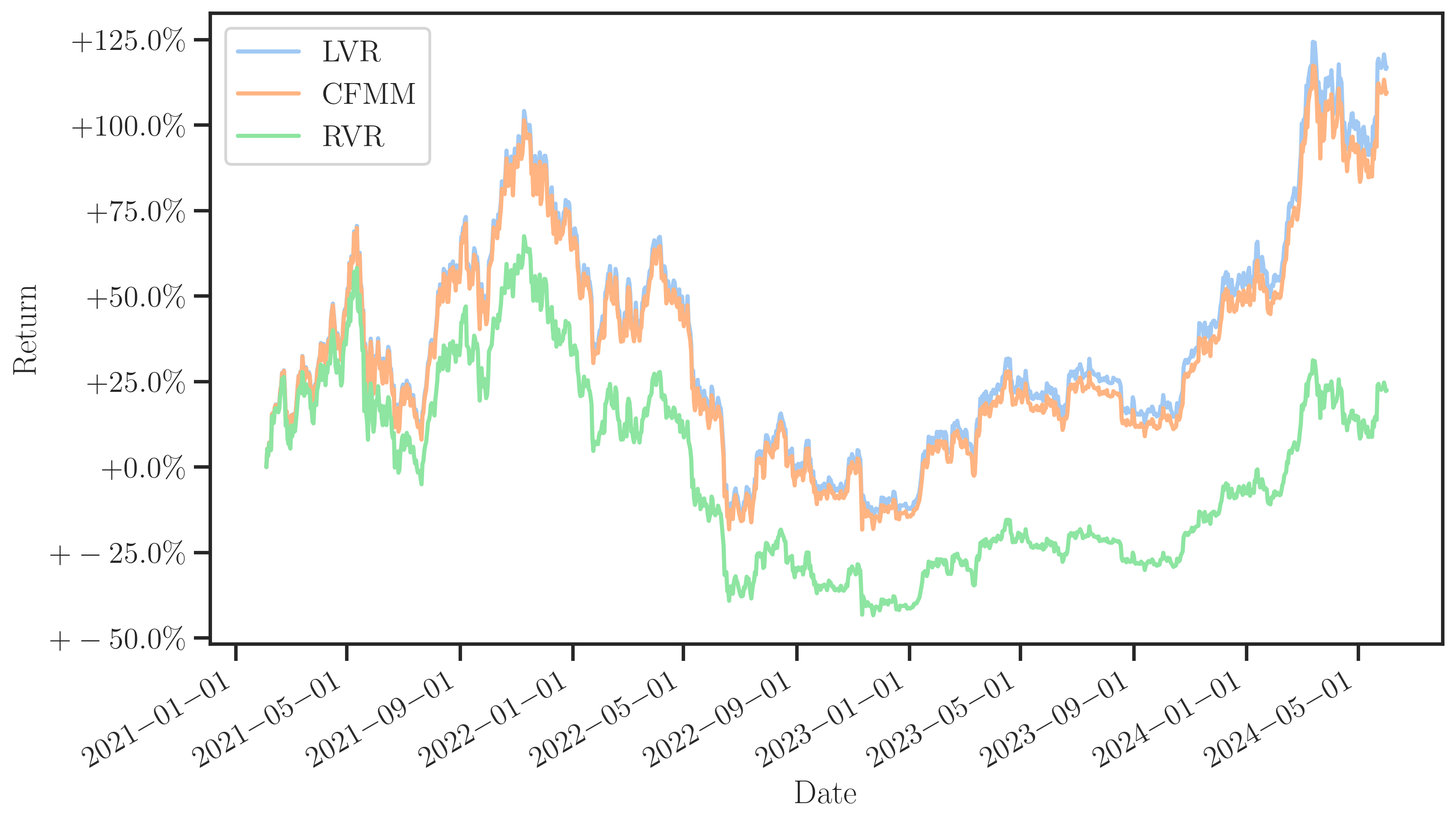}            \caption[]%
            {{\small RVR and LVR for a CFMM (constant) pool}}    
            \label{fig:cfmm_lvr_rvr}
        \end{subfigure}
\begin{subfigure}[b]{0.45\textwidth}
            \centering
        \includegraphics[width=\textwidth]{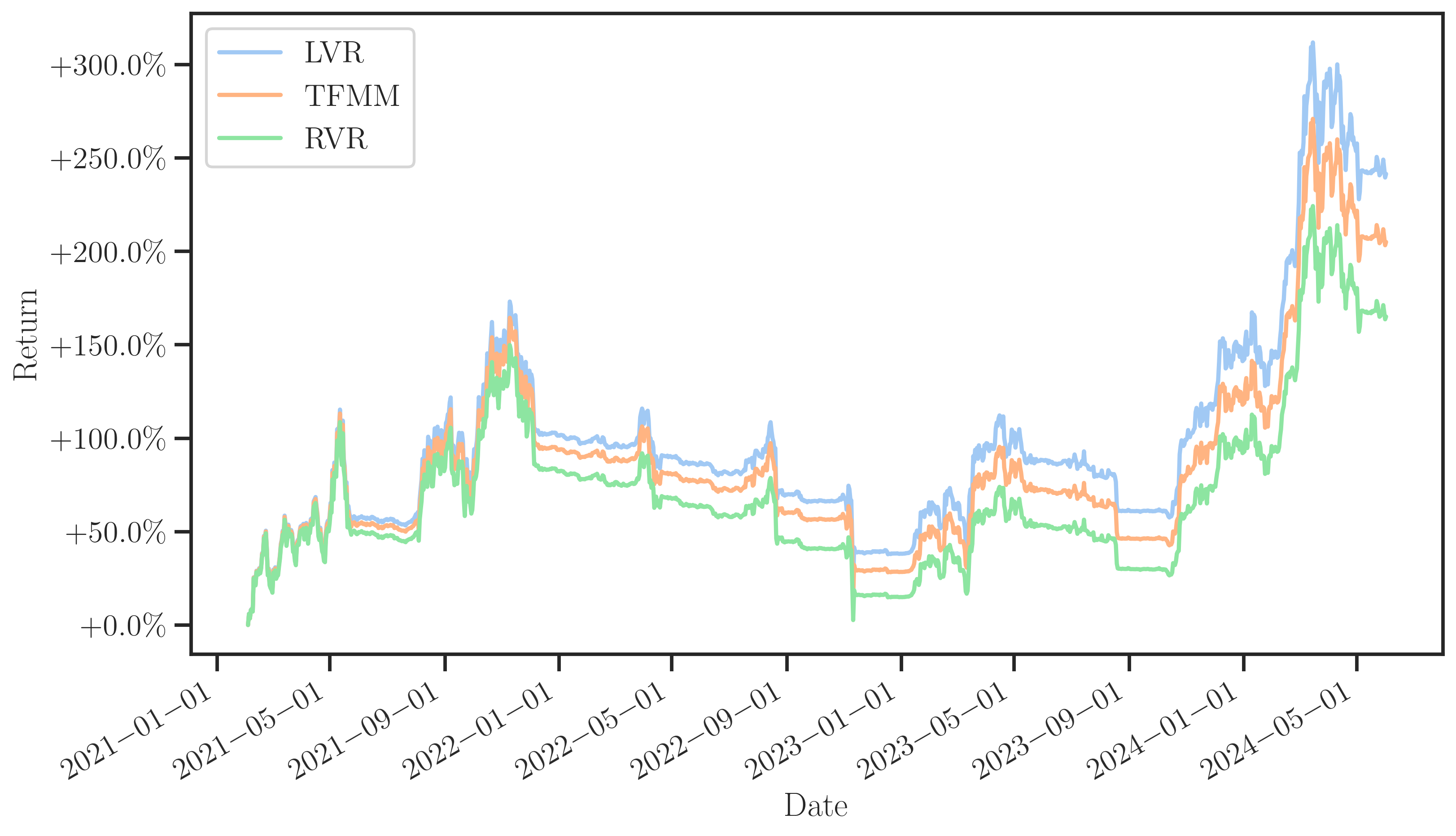}
            \caption[]%
            {{\small RVR and LVR for a TFMM (dynamic) pool}}    
            \label{fig:tfmm_lvr_rvr}
        \end{subfigure}
        \hspace{3em}
        
     \caption{Cumulative returns for a) a TFMM pool (with dynamic weights from a momentum strategy) and b) a CFMM pool (constant weights). The pools contain BTC, ETH and DAI. Each is shown with the benchmark values achieved by rebalancing under LVR (blue) and RVR (green). RVR rebalancing is calculated for a CEX with 20 bp fees. Note that pools (orange) both beat RVR here, indicating that implementing these strategies as AMM pools may be superior to running them by trading with a CEX.}
     \label{fig:rvr_intro}
\end{figure}

In Fig~\ref{fig:rvr_intro} we give an example of the results of simulations showing both TFMM (dynamic) and CFMM (constant weight) pools beating an RVR benchmark; this finer-grained modelling indicates that running strategies as AMM pools rather than directly rebalancing them on a real-world CEX could provide better performance.

\paragraph{Rebalancing Efficiency vs Impermanent Loss}

It is well known that first generation CFMMs, more often than not, generated losses for LPs in the long run over the counterfactual of LPs simply holding their initial assets~\cite{loesch2021impermanentlossuniswapv3}.
How can this be reconciled against beating RVR benchmarks even for fixed weights? 

There is a distinct difference between rebalancing efficiency and a good strategy. Given that CFMMs beat their RVR benchmark, it can be determined that CFMMs are more efficient for maintaining fixed weights given the rebalancing frequency and historical price volatility--as well as all the other settings. 

Does this mean that fixed weights are a good overall strategy for LPs? Absolutely not. LVR and RVR do not benchmark against HODL, as in Impermanent Loss: they do not take into account the relative market value of a pool against other possible strategies.
Performing a bad asset management strategy as efficiently as possible does not make it a good asset management strategy. 

\newpage
\subsection{Results varying rebalancing parameters}

\begin{figure}[H]
\centering
\includegraphics[width=0.55\textwidth]{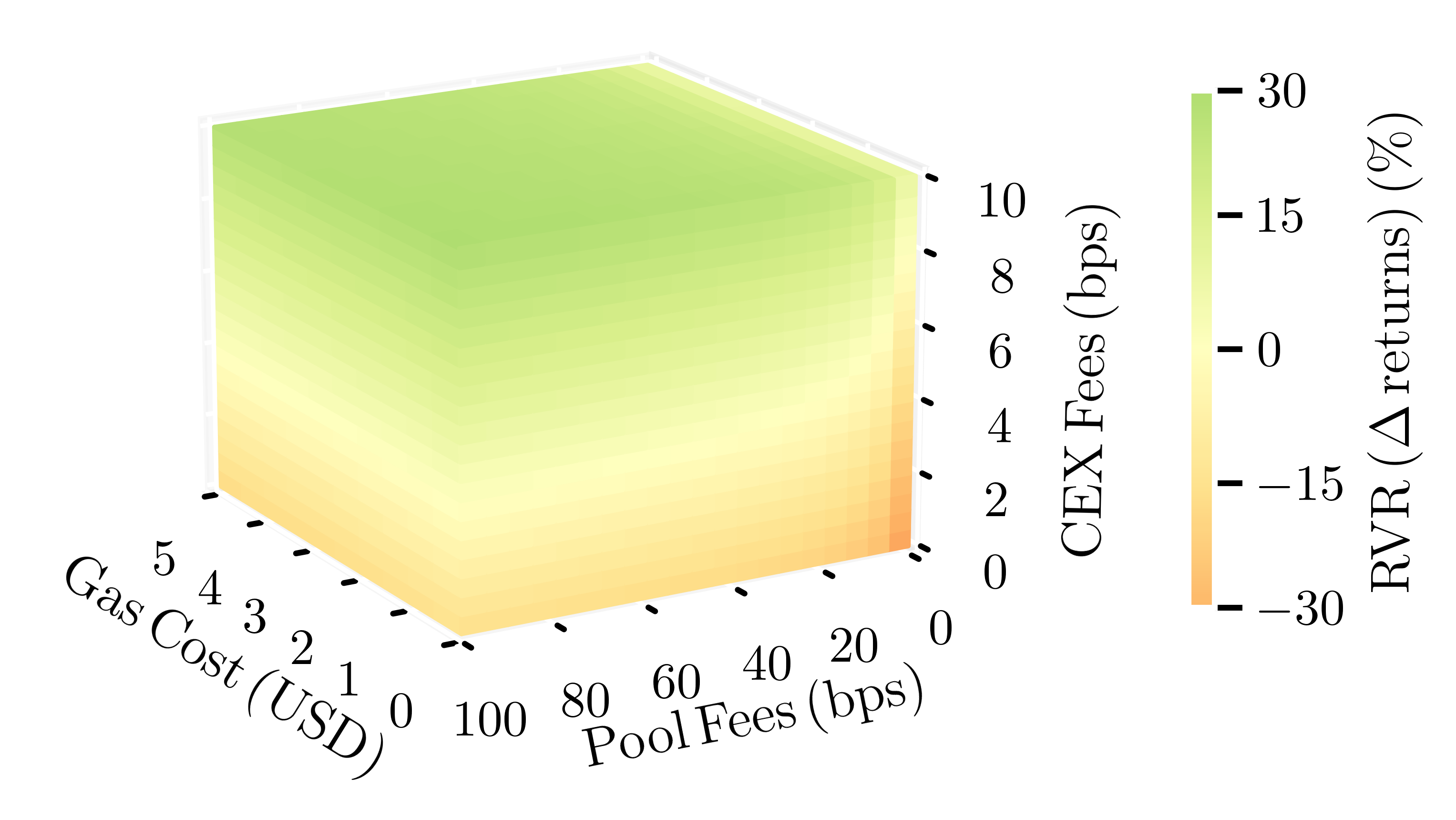}
            \caption[]%
            {{A 3D heatmap showing RVR for variations in AMM fees, gas cost and CEX fees. As expected, RVR increases for low gas costs and high CEX fees. Further, increasing pool fees tends to increase performance. The pool has an initial TVL of \$50m }}
            \label{fig:rvr_cube_no_noise}
\end{figure}

It should be of no surprise that AMM execution costs increasing by either increasing gas cost or fees reduces the relative performance of TFMMs vs the RVR benchmark.
In Fig~\ref{fig:rvr_cube_no_noise} we study one dynamic strategy for a broad range of different rebalancing parameters: a useful range of CEX fees that one can expect as a VIP price taker as well as appropriate ranges for AMM settings.
For the individual plots that are combined here, see Fig~\ref{fig:rvr_fee_heatmaps}.

At the extreme VIP CEX fee volumes the reduced cost of fees means that CEX rebalancing efficiency starts to be optimal. A larger exploration of parameters is available  in [cite]. Usual monthly volumes required to reach these leaves are in the high tens of millions and given previous observations that monthly volumes are close to allocation, a strategy allocation in the high tens of millions. 

\subsubsection{Exploring AMM noise trading}

All results so far have been extremely conservative with the AMM receiving no retail swap fee income. 
$\sim 40\%$ of trade volume on Uniswap V3 is from arbitrage \nbcite[\S5.3]{canidio2024arbitrageursprofitslvrsandwich}, \cite{heimbach2024nonatomicarbitragedecentralizedfinance}, $\sim 30\%$ being non-atomic arbitrage and $\sim6\%$ from sandwich attacks and atomic arbitrage, meaning $\sim60\%$ of live trade volumes can be attributed to noise traders.
This means that noise traders are responsible for about $1.5\times$ the volume that arbitrage traders produce (60\% is $1.5\times40\%$).
We choose to be conservative and introduce a deliberately low amount of noise trade volume, just $1\times$ arbitrage volume.
This may be extremely conservative, given that dynamic AMMs offer off-market prices to incentivize arbitrage that may be advantageous for noise traders increasing volumes above that of vanilla AMMs.

\begin{figure}[H]
\centering
\includegraphics[width=0.55\textwidth]{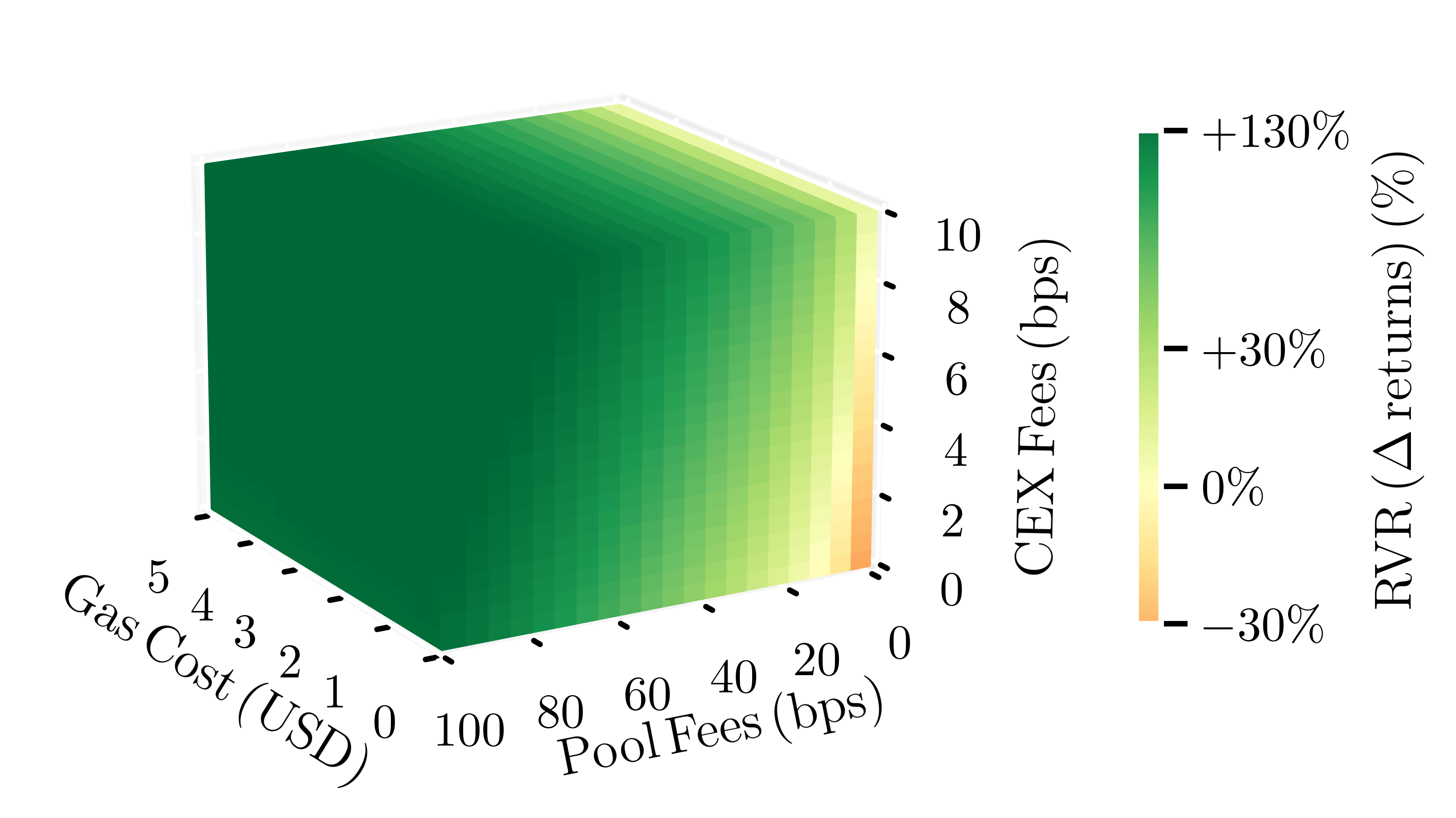}
            \caption[]%
            {{A 3D heatmap showing RVR for variations in AMM fees, gas cost and CEX fees. Initial TVL is still kept at \$50 million. Note that the colours for the cube surface are the same scale as in figure.}}
            \label{fig:rvr_cube_noise}
\end{figure}

\section{Conclusion}

\subsection{Summary}
Improving the realism of the LVR benchmark demonstrates that, prior to any retail swap fee income, running a portfolio as a dynamically weighted AMM, allowing external arbitrageurs to rebalance for you, is not only viable asset management approach but often a superior one in our modelling framework.
Introducing even low levels of noise trading demonstrates a rebalancing efficiency that even the highest volume-discount CEX fee levels can not achieve. 

For this reason we believe RVR can become an important benchmark for direct comparisons and decision making of capital allocators and, given that it can be beaten, standard TradFi metrics and analysis applying RVR as a benchmark return $R_b$ can be performed. 

For single strategy funds and accounts, over practical strategy variations tested and sensible AMM parameters, the results in this paper show that dynamic AMMs can offer superior execution efficiency for all but >\$100m monthly volume CEX fee bands that offer less that 2bps fees.
Applying low/realistic levels of retail swap fee income may mean that dynamic AMMs become superior to even 0 fee CEX structures. 

\subsection{Limitations and potential expansions of RVR}

While there does seem to be a natural limit to the benefits given the parameters of the above tests, the AMM modelling is handicapped to liner weight interpolation methods: 

Non-linear methods for weight trajectories have been shown to provide further efficiency~\cite{willetts2024optimalrebalancingdynamicamms}.
Those advances are not used in this analysis, and indeed the current work on optimal trajectories assumes zero pool fees.
Weight trajectories could be developed that aim to improve a pool's RVR.

Retail swap fee revenue modeling is a complex and subjective area of study.
This was excluded in all but the final results to simplify the initial experiments, while making it harder for AMMs to beat RVR.
Even in our exploration, noise trading effects on arbitrage volumes and off-market prices effects have not been included, only the simple fee income.

In RVR modelling an extension which may become more relevant at extrememly large portfolio sizes is market impact.
In some initial exploratory work we have found that, as TFMM strategies naturally reduce market impact already, only very small changes come from including market impact to RVR, but for other strategies this might be a more important aspect to include.

Discussion so far has been limited to single strategy pools and porfolios.
The implications of RVR, and of utilising dynamic AMMs as execution infrastructure, are not limited to single strategy funds/pools. Multi-strategy portfolios can be implemented as dynamic AMMs given composite pool structures both currently available and expanded upon in the TFMM papers~\cite{tfmm_litepaper, tfmm}.

\newpage
\bibliography{biblio}

\newpage
\begin{appendices}
\renewcommand\thefigure{\thesection.\arabic{figure}}    
\renewcommand\theequation{\thesection.\arabic{equation}}    
\section{Supplementary figures}
\subsection{Heatmaps of RVR for 1600 strategies, with varying gas cost and pool fees}
\begin{figure*}[h]
        \centering
        \begin{subfigure}[b]{0.475\textwidth}
            \centering
            \includegraphics[height=0.8\textwidth]{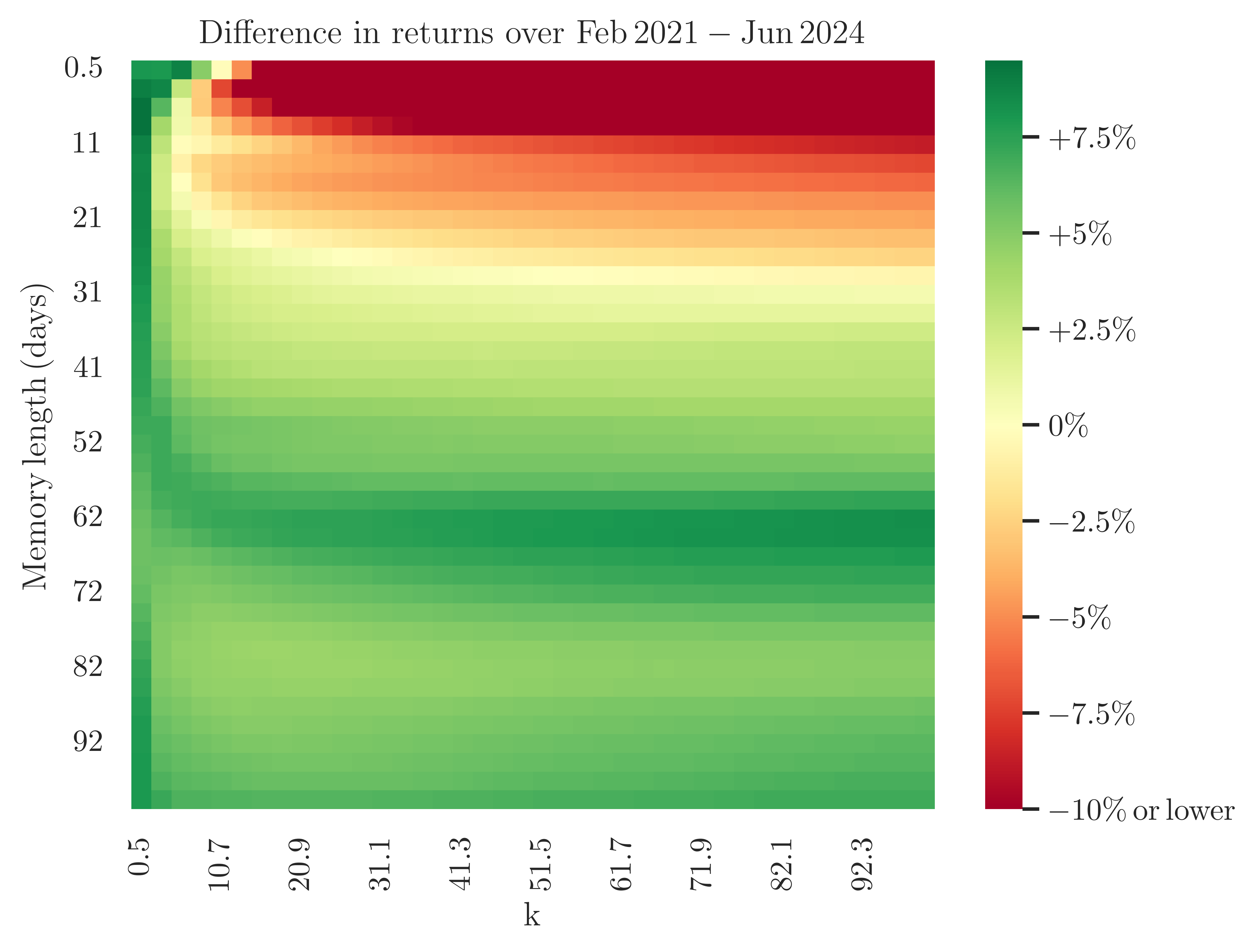}
            \caption[]%
            {{\small $\tau_\mathsf{cex}=0$, gas cost = $1$ USD}}    
            \label{fig:fee0gas1}
        \end{subfigure}
        \hfill
        \begin{subfigure}[b]{0.475\textwidth}  
            \centering 
            \includegraphics[height=0.8\textwidth]{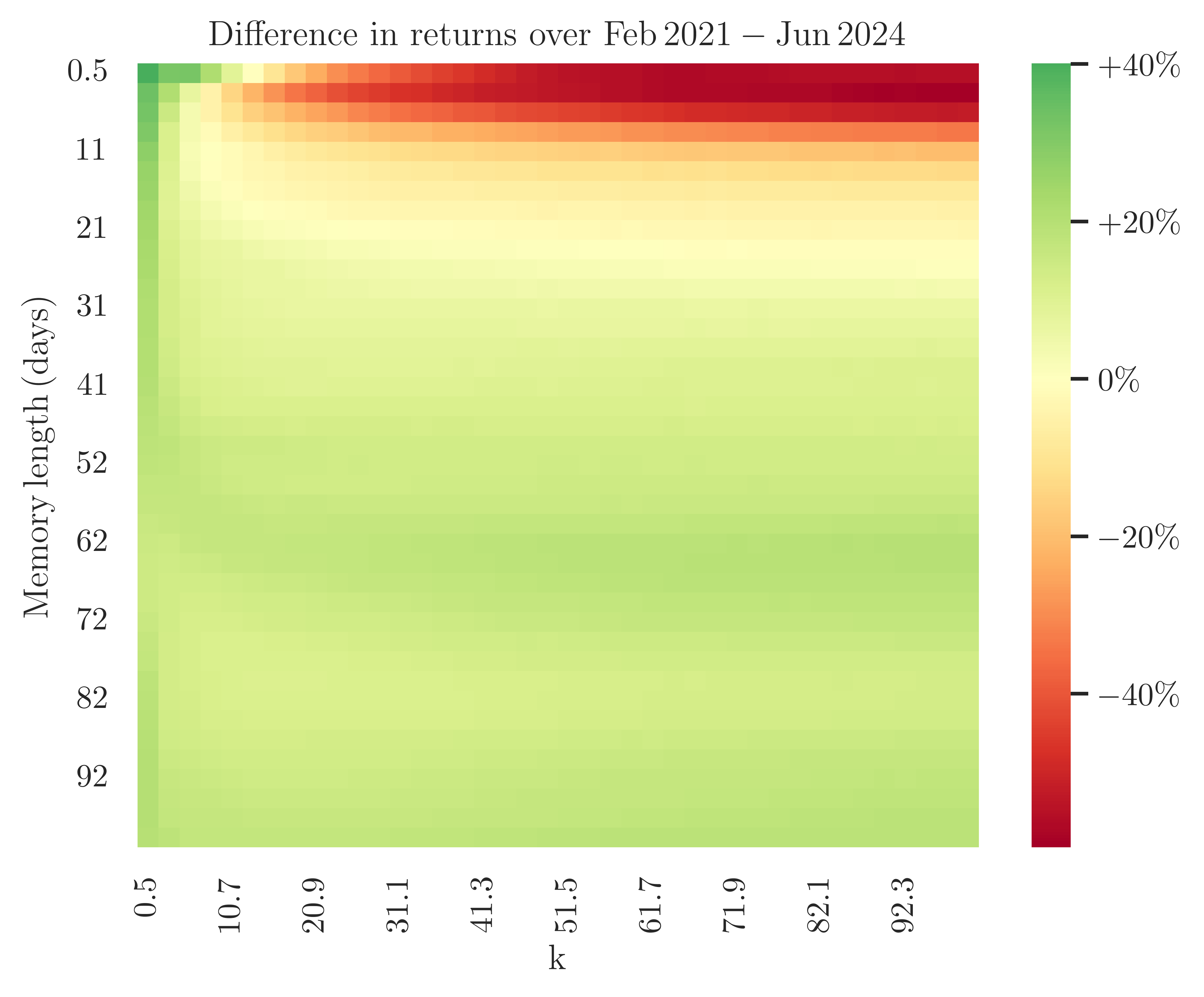}
            \caption[]%
            {{\small $\tau_\mathsf{cex}=0.014$, gas cost = $1$ USD}}    
            \label{fig:fee1gas1}
        \end{subfigure}
        \vskip\baselineskip
        \begin{subfigure}[b]{0.475\textwidth}   
            \centering 
            \includegraphics[height=0.8\textwidth]{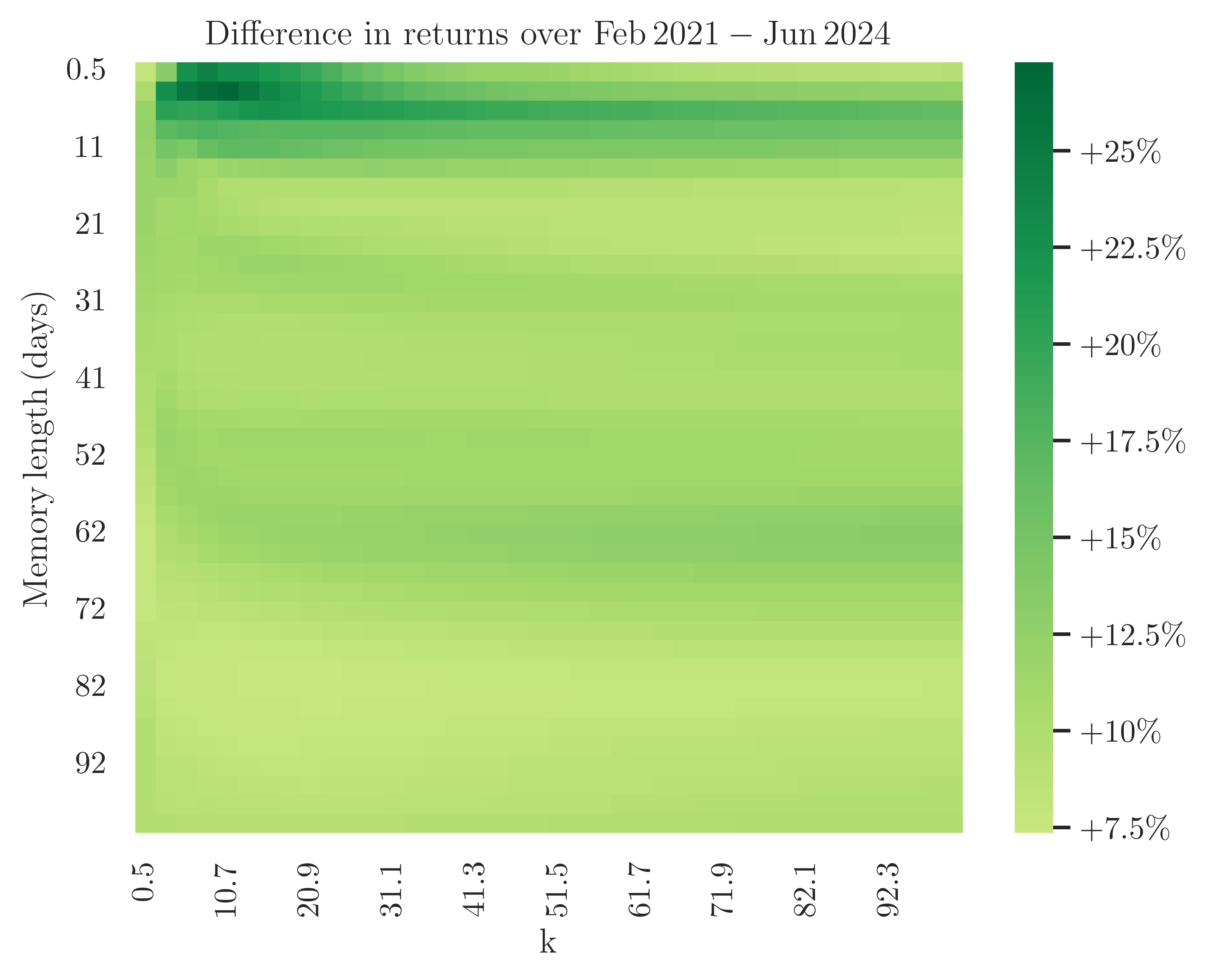}
            \caption[]%
            {{\small $\tau_\mathsf{cex}=0$, gas cost = $0$ USD}}    
            \label{fig:fee0gas0}
        \end{subfigure}
        \hfill
        \begin{subfigure}[b]{0.475\textwidth}   
            \centering 
            \includegraphics[height=0.8\textwidth]{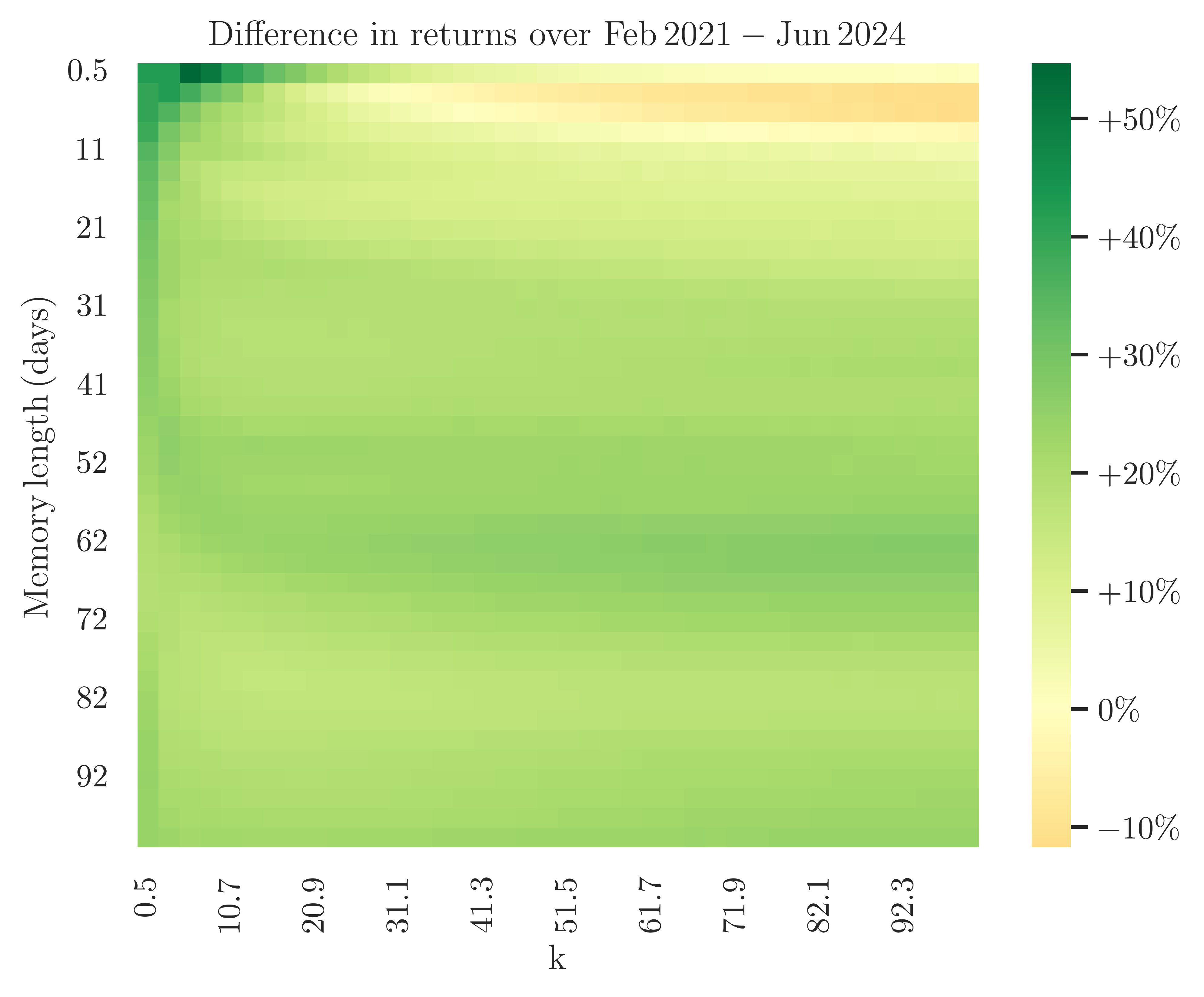}
            \caption[]%
            {{\small $\tau_\mathsf{cex}=0.014$, gas cost = $0$ USD}}    
            \label{fig:fee1gas0}
        \end{subfigure}
\caption{RVR plotted for 1600 different momentum strategies, for different gas costs and pool fes. Note that increasing gas cost, all else equal, increases reduces RVR while increasing pool fees increases RVR.}
     \label{fig:backtest}
    \end{figure*}

\newpage
\subsection{Heatmaps of RVR, varying gas cost, pool fees and CEX fees}
\begin{figure}[h]
\centering
\begin{subfigure}[b]{0.45\textwidth}
            \centering
\includegraphics[width=\textwidth]{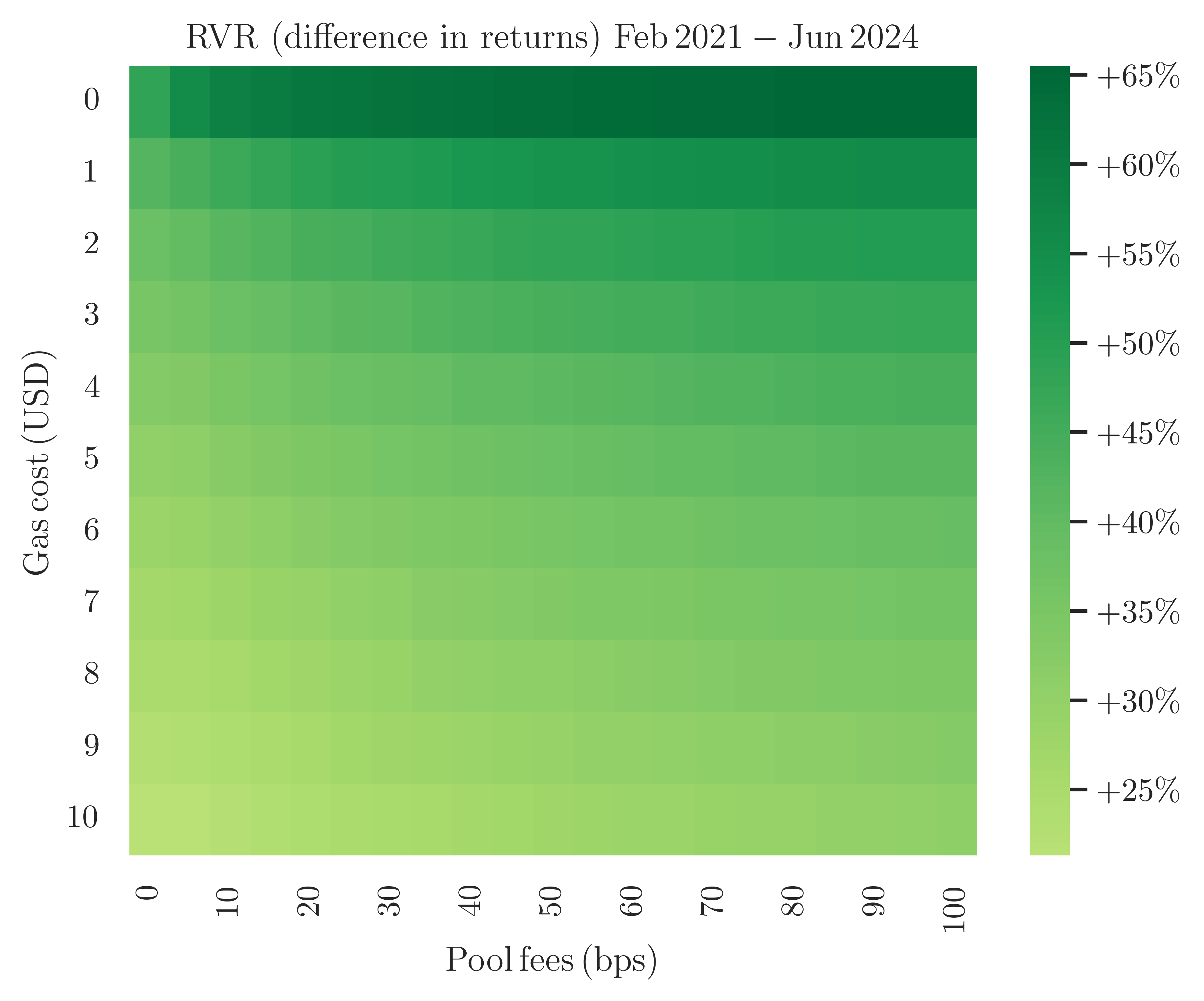}
\caption[]%
{{\small RVR when varying pool fees and gas cost. RVR here is calculated with CEX commission fess $\tau_\mathsf{cex}=25$ bps.}}    \label{fig:tfmm_param_rvr_cex_fees_constant}
\end{subfigure}
\begin{subfigure}[b]{0.45\textwidth}
\centering
    \includegraphics[width=\textwidth]{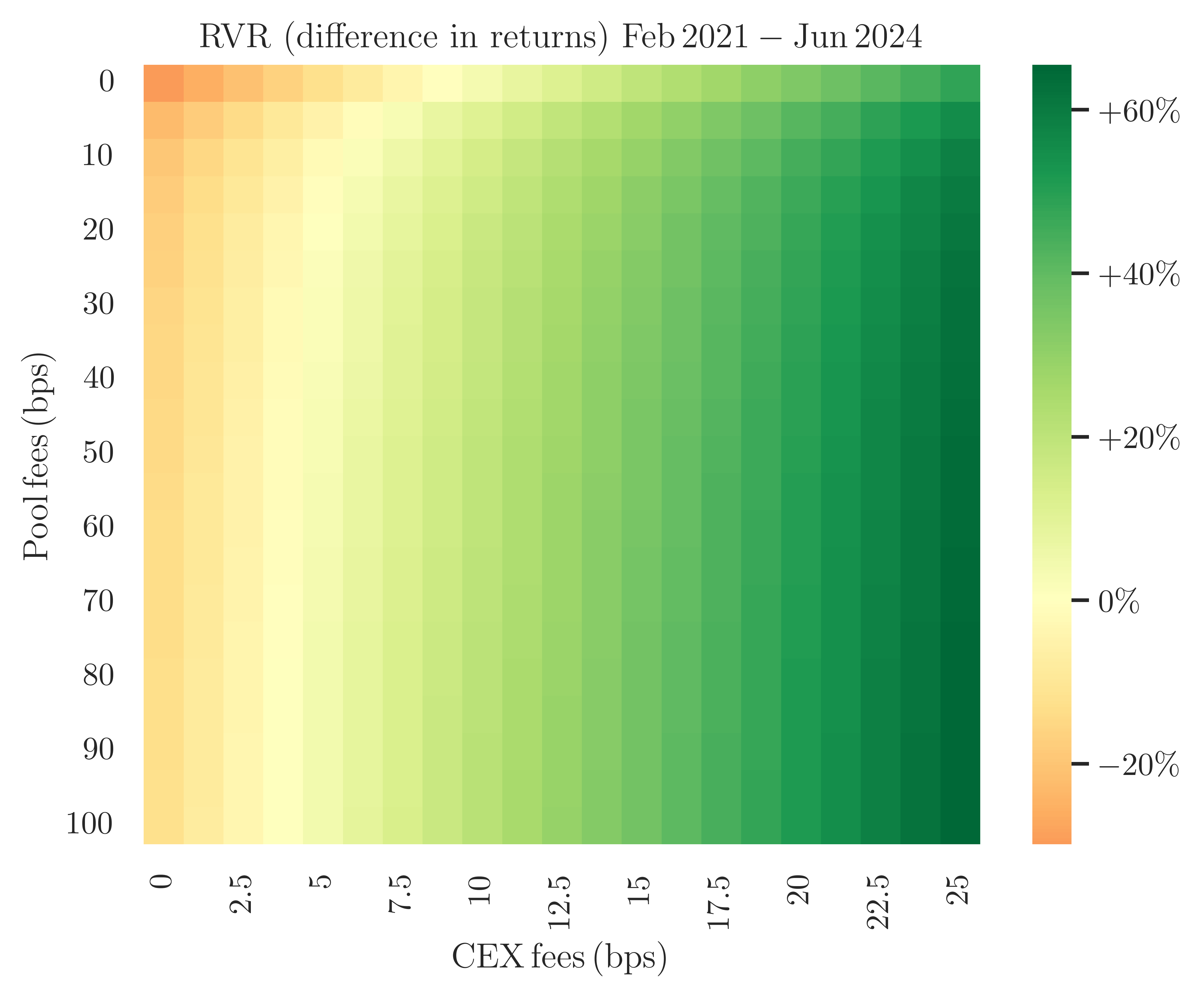}
\caption[]%
            {{\small RVR when varying pool fees and CEX commission fees. RVR here is calculated with gas cost of $1$ USD.}}    
            \label{fig:tfmm_param_rvr_gas_fees constant}
\end{subfigure}
        \vspace{1em}
\begin{subfigure}[b]{0.45\textwidth}
\centering
\includegraphics[width=\textwidth]{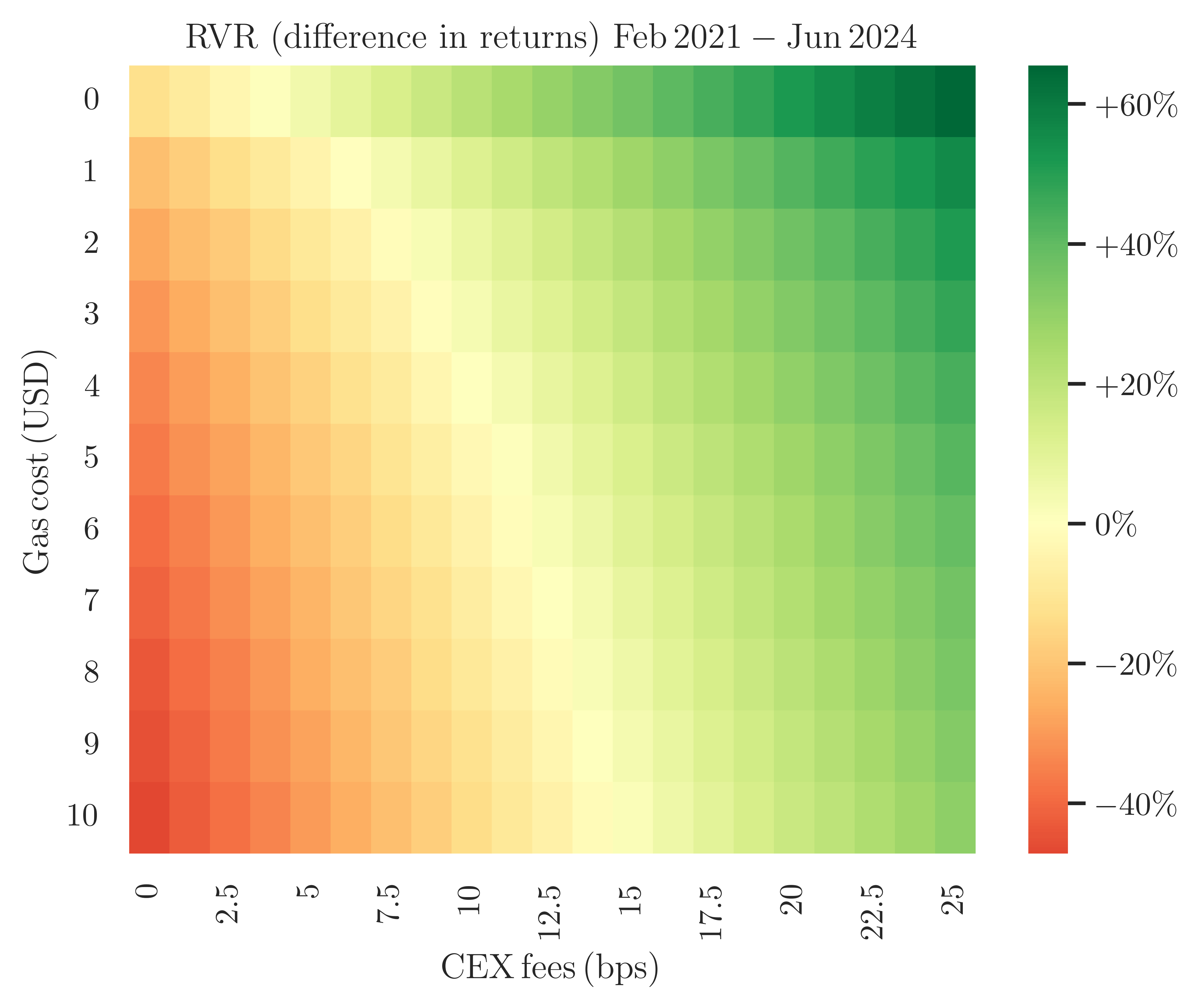}

\caption[]%
            {{\small RVR when varying trade gas cost and CEX commission fees. RVR here is calculated with pool fees of $100$ bps ($\gamma=0.99$).}}
            \label{fig:tfmm_param_rvr_pool_fees}
\end{subfigure}
\caption[Network2]%
            {{\small $\frac{\mathsf{RVR}(t)}{V(0)}$, RVR scaled by initial value, variations of different pool fees (21 values from 0 to 100 bps), trade gas costs (11 values from 0 to 10 USD) and CEX fees (21 values from 0 to 25 bps).}}
            \label{fig:rvr_fee_heatmaps}
\end{figure}
\newpage
\section{Multi-Asset generalisation in Original LVR paper}
\label{app:multi_asset_lvr}
In \nbcite[App.B.3]{lvr} the authors propose a multi-asset generalisation of LVR.
Like us, they aim to have a version of LVR that can naturally extend beyond pairs and removes the risk-on/risk-off lens for viewing assets.

Recall that in 2-asset LVR, the rebalancing strategy matches the holdings of the AMM pool's holdings \emph{only for the risk asset}, meaning that the rebalancing strategy can have different (always equal or greater) holdings of the risk-off asset than the AMM pool.
Indeed, it is \emph{from} this difference in holdings that the rebalancing strategy can have a different (always equal or greater) value than the AMM pool.

Using notation from~\cite{lvr}, to get to their multi-asset generalisation introduce an $N$-vector of reserves $\v x$ and an $N$-vector of market prices $\v P$.

The value of a pool is thus $V=\v x^\top \v P$ and $\v x^*(\v P) = \nabla V(\v P)$ (see Lemma 2 in \nbcite[B.3]{lvr}).
They define the rebalancing portfolio's value to be $V_0+\int_0^t \v x^*(\v P_s)^\top \mathrm{d}\v P_s$ for $t>0$ and where $\v x_t=\v x^*(\v P_t)$ are the holdings of the pool.

This means that under their definition the rebalancing strategy identically matches the holdings of the AMM pool \emph{exactly and for all assets}, meaning that the value of the AMM pool's holdings and the rebalancing strategy's holdings are also identical, rendering their multi-asset generalisation unhelpful.
\end{appendices}

\newpage
\textbf{DISCLAIMER} This paper is for general information purposes only.
It does not constitute investment advice or a recommendation or solicitation to buy or sell any investment or asset, or participate in systems that use TFMM.
This paper should not be used in the evaluation of the merits of making any investment decision.
It should not be relied upon for accounting, legal or tax advice or investment recommendations.
This paper reflects current opinions of the authors regarding the development and functionality of TFMM and is subject to change without notice or update.

While some aspects, such as altering target weights in geometric mean market makers is prior art, aspects of TFMM that are novel such as, but not exclusively, composability mechanisms, efficient methods for gradients and covariances, generic form multi-token amplification and advanced execution management mechanisms for use in dynamic weight AMMs, for purposes of core liquidity providing or forms of asset management including, but not exclusively, fund construction, structured products, treasury management are covered by patent filing date of 21st February 2023.

\end{document}